Revised Paper:Theory of opinion distribution in human relations where trust and distrust mixed(2020)

# Massive Case Study of Opinion Distribution in a Relationship with Mixed Trust and Distrust

Yasuko Kawahata [†]

Faculty of Sociology, Department of Media Sociology, Rikkyo University, 3-34-1 Nishi-Ikebukuro,Toshima-ku, Tokyo, 171-8501, JAPAN.
ykawahata@rikkyo.ac.jp,kawahata.lab3@damp.tottori-u.ac.jp

**Abstract:** The simulations in this paper are based on the theory of opinion dynamics, which incorporates both Opinion A and Opinion B, a case that is the inverse of Opinion A, in human relationships. It was confirmed that aspects of consensus building depend on the ratio of the trust coefficient to the distrust coefficient. In this study, the ratio of trust to distrust tended to vary like a phase transition around 55%, but we wanted to see if the same phenomenon could be confirmed in large-scale cases. In the previous case studies, this tendency has been observed from $N = 300$ to $N = 1600$, and we will discuss the case of $N = 10000$ with $N = 3000$. By verifying the extent of the phenomenon on the social scale, we intend to consider simulation items for consensus building, such as consideration of the sensitivities of topics in online and offline opinion formation.

**Keywords:** Opinion Dynamics, Trust-Distrust Model, Socio-Dynamics, Large-scale Simulation

## 1. Introduction

With the proliferation of public networks, public networking devices are spreading across the globe. As a result, opportunities for decision-making and consensus building beyond spatio-temporal constraints have increased significantly. In other words, online is playing a major role as a concrete physical place for disagreement and consensus building. In the future, it is anticipated that simulation discussions on opinion formation among many people will be necessary. The issue of consensus building through social exchange has been studied for a long time. In many cases, however, the theory is primarily an analytical approach based on consensus building. Once a consensus is reached, people's opinions tended to eventually converge to a small number, as the Trusted Model shows. However, the distribution of opinions in the real world cannot be summed up into a small number of opinions. At least, we assume that they cannot be summed up, depending on the results of the simulation. Ishii and Kawahata et al. proposed a theory of opinion dynamics that includes both trust and distrust relationships in human networks in society. In this paper, the theory is applied to the distribution of opinions in a society that includes both trust and distrust. The paper discusses the tendency for phase transition-like behavior that depends on the parameter of the ratio of trust to distrust in relationships in a society. The simulations in this paper are based on the theory of opinion dynamics, which incorporates both Opinion A and Opinion B, a case that is the inverse of Opinion A, in human relationships. It was confirmed that aspects of consensus building depend on the ratio of the trust coefficient to the distrust coefficient. In this study, the ratio of trust to distrust tended to change like a phase transition around roughly 55% in existing studies, and we wanted to confirm whether the same phenomenon is confirmed in large-scale cases. In previous case studies, this tendency has been observed from about $N = 300$ to $N = 1600$. In this paper, we discuss the case of $N = 10000$ with $N = 3000$, because the expansion of computing resources, which has been an issue until now, has made it possible to increase the speed of processing. By verifying the extent of the phenomenon on the social scale, we intend to consider simulation items for consensus building, such as consideration of the sensitivities of topics in online and offline opinion formation.

## 2. Trust-Distrust Model, Opinion Distribution

Opinion dynamics theory is applied to compute simulations of human behavior in society. In this paper, we introduce distrust into the bounded trust model in order to discuss the time transition and trust between the two. For a fixed agent, $1 \leq i \leq N$, the agent's opinion at time $t$ is $I_i(t)$. As a trust coefficient, we modified the meaning of the coefficient $D_{ij}$ in the bounded trust model. Here, we assumed that $D_{ij} > 0$ if there is trust between them and $D_{ij} < 0$ if there is distrust between them. For the calculations in this paper, $D_{ij}$ was assumed to be constant. Thus, the change in the opinion of agent $i$ can be expressed as follows:

$$\Delta I_i(t) = -\alpha I_i(t) + c_i A(t)\Delta t + \sum_{j=1}^{N} D_{ij} I_j(t)\Delta t \quad (1)$$

$$D_{ij}\phi(I_i, I_j)(I_j(t) - I_i(t)) \quad (2)$$

$$\phi(I_i, I_j) = \frac{1}{1 + \exp(\beta(|I_i - I_j| - b))} \quad (3)$$

Here, in order to cut off the influence from people whose opinions differ significantly(3), we use the following sigmoidal smooth cutoff function, which is a Fermi function system. In other words, the model hypothesizes that people do not pay attention to opinions that are far from their own. We will have a separate discussion on the introduction regarding this Fermi function system in the future. Here, $D_{ij}$ and $D_{ji}$ are assumed to be independent. Usually, $D_{ij}$ is an asymmetric matrix. In the following calculations, $a = 1$ and $b = 5$. Furthermore, it is assumed that $D_{ij}$ and $D_{ji}$ can have different signs. In the actual calculations below, we present simulation calculations using Ishii theory for 300 persons. For simplicity, we assume that the 300 links form a complete graph. $D_{ij}$ for each link is assumed to be a value between -1 and 1 using random numbers in the actual calculation. In the actual calculation, the ratio of positive to negative values is set differently when $D_{ij}$ takes values between 0 and 1 and the ratio of values between -1 and 0 is set to 0. In the case of the previous computational experiments, the initial opinion values for $N = 300$ people are randomly determined between -20 and +20. As a result of the opinion dynamic model calculation, the trajectories of the 300 opinions are calculated and their histograms are simulated as opinion distributions. In this paper, we discuss the results for $N = 10000$ with $N = 3000$, improving on a problem of computational resources, and we discuss whether the phase transition phenomenon at the 55% threshold confirmed in the previous cases can be confirmed for very large numbers of people, and if not, whether the threshold for the phase transition so far can be approximated. If not, we would like to examine whether or not the opinion model itself needs to be examined in light of other environmental variables such as external factors, conditions, magnetic field conditions (invariant or variable), etc.

## 3. Discussion

**Mathematical Formulation for Large Scale $D_{ij}$ <0.1**

(1) **Opinion Update Equation**:
The update of each individual's opinion is based on the following equation:

$$o_i^{(t+1)} = o_i^{(t)} + h\left(a + \sum_{j\neq i}^{n} d_{ij}(o_j^{(t)} - o_i^{(t)}) \times \Phi(o^{(t)}, \beta, \alpha, j, i) - d_{ii}o_i^{(t)}\right) \quad (4)$$

Where:

$o_i^{(t)}$ is the opinion of individual $i$ at time $t$.

$h$ is the time step interval.

$a$ is a constant scalar.

$d_{ij}$ represents the degree of trust (or distrust) individual $i$ has for individual $j$.

(2) **Phi Function**:
This function returns a value between 0 and 1 based on the difference between two opinions:

$$\Phi(u, \beta, \alpha, j, i) = \frac{1}{1 + \exp(\beta(|u_j - u_i| - \alpha))} \quad (5)$$

Where:

$u$ is the vector of opinions.

$\beta$ and $\alpha$ are constant parameters.

This equation is designed to return a value close to 0 when the difference between two opinions is large and a value close to 1 when the difference is small.

(3) **Initialization of the Trust Matrix**:

$$d_{ij} = \begin{cases} \text{random value between 0 and 1} & \text{if } x < 0.1 \\ \text{negative random value between -1 and 0} & \text{otherwise} \end{cases} \quad (6)$$

The "Opinion dynamics over time" graph shows the movement of opinion over time. It is observed that the fluctuation of opinions varies greatly depending on the time of day,

The "Final distribution of opinions" graph shows the final distribution of opinions. The graph here shows that the distribution of opinions has a peak near the center, but is spread over a wide range.

It appears to generate a within-group and between-group confidence coefficient $Dij$, which is set to take a positive random value if a random value smaller than 0.8 is generated, and a negative random value otherwise.

Based on this information, let us consider the following two points.

Figure 1 and Figure 2,

(1) **Consideration as a process of consensus building under conditions of trust and distrust**

(a) **Existence of trust**

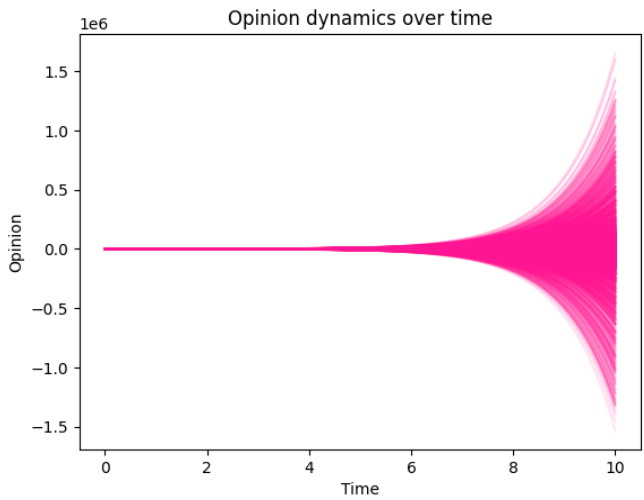

Fig. 1: Calculation result for Opinion dynamics over time $D_{ij}<0.1$, $N = 10000$

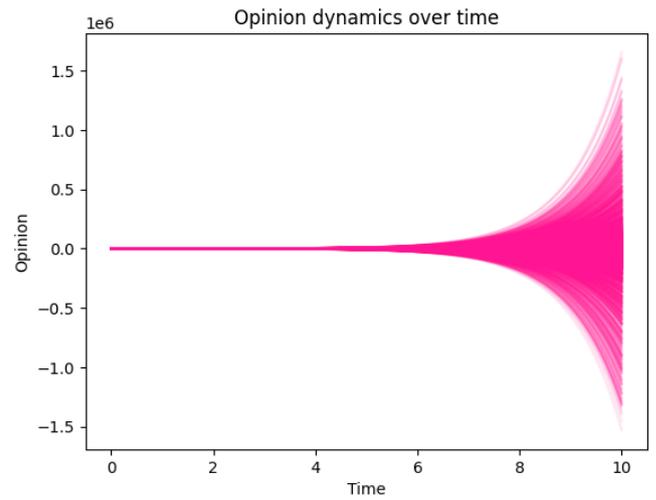

Fig. 3: Calculation result for Opinion dynamics over time $D_{ij}<0.1$, $N = 3000$

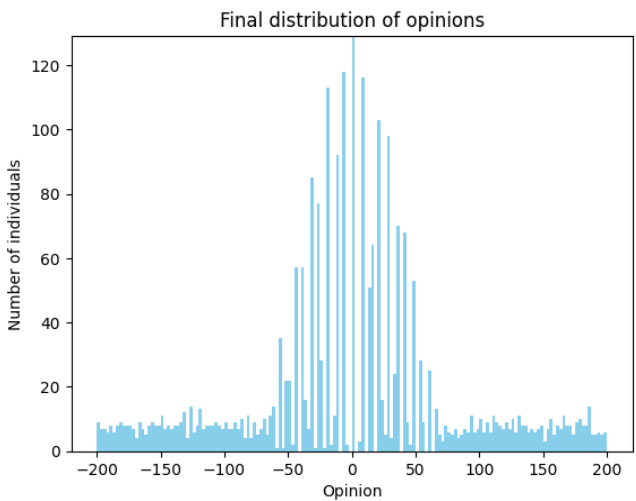

Fig. 2: Calculation result for Final distribution of opinions $D_{ij}<0.1$, $N = 10000$

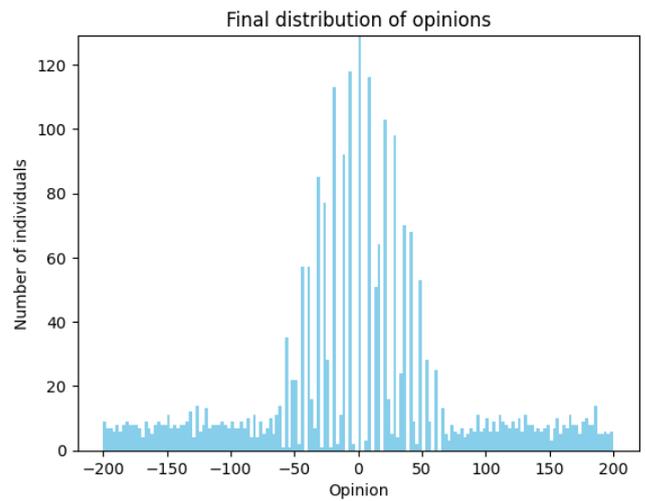

Fig. 4: Calculation result for Final distribution of opinions $D_{ij}<0.1$, $N = 3000$

- *Center of agreement:* In the "Final distribution of opinions" graph, we see that many individuals have opinions in the neighborhood of 0. This indicates that many people share a near-neutral opinion, which can be interpreted as a characteristic of the presence of trust.
- *Less dispersion:* In this model, the distribution of opinions can be narrower when trust relationships are formed. This means that people are more likely to be influenced by the opinions of others they trust, thus increasing the likelihood of convergence of opinions.

(b) **Presence of distrust**

- *Opinion dispersion:* On the other hand, we can also observe from the graph that the distribution of opinions is spread out. This can be attributed to the presence of distrust. People in distrustful relationships are less likely to be influenced by the opinions of others, and therefore may hold to their own opinions.
- *Extreme opinions:* The "Opinion dynamics over time" graph shows that some people become more extreme in their opinions as time goes by. This may be due to the presence of distrust, which leads some people to extreme opinions.

(2) **What is the case in society, consideration of opinion formation**

*Information bias:* In society, bias in media and information sources can have a significant impact on opinion formation. The results of the above model suggest that some opinions tend to become mainstream and those that do not follow them tend to be extreme. This phenomenon can be related to issues in contemporary society such as media bias and the echo chamber phenomenon.

(3) **Power of Social Influence**

This model shows how much an individual's initial opinions and trust relationships influence subsequent opinion formation. When trust is formed, people are more likely to be influenced by the opinions of others, resulting in a greater tendency for opinions to converge. This indicates that the power of social influence plays a major role in the formation of individual opinions.

(5) **Insight**

The above discussion shows that the balance between trust and distrust is very important in the opinion formation process. Many factors are involved in the formation of opinions in society, among which the existence of trust plays a central role.

Figure 3 and Figure 4, comparison with the aforementioned $N = 10000$ cases,

As for the effect of sample size, the $N = 3000$ case may have a different speed of opinion variation and convergence than the N=10000 case due to the smaller sample size. With a smaller sample size, the impact of individual opinions may be greater. In terms of the distribution of confidence, based on the code provided, the settings are such that 10% of the time the confidence level takes a positive value (confidence) and 90% of the time the confidence level takes a negative value (distrust). Under such a distribution, one would expect slower convergence of opinions due to the high number of distrust situations; a direct comparison with the data from the N=10000 case is needed to analyze in detail whether this trend is stronger or weaker than in the $N = 100000$ case. As for the consensus building process, In the case of $N = 3000$, there is also a trend toward greater initial variation in opinion and decreasing variation in opinion over time. This is a similar trend to the $N = 10000$ case. However, it is possible that there may be differences in the specific values and shapes of the final distribution of opinions and the speed of convergence.

**Mathematical Formulation for Large Scale $D_{ij}$<0.5**

Figure 5 and Figure 6,

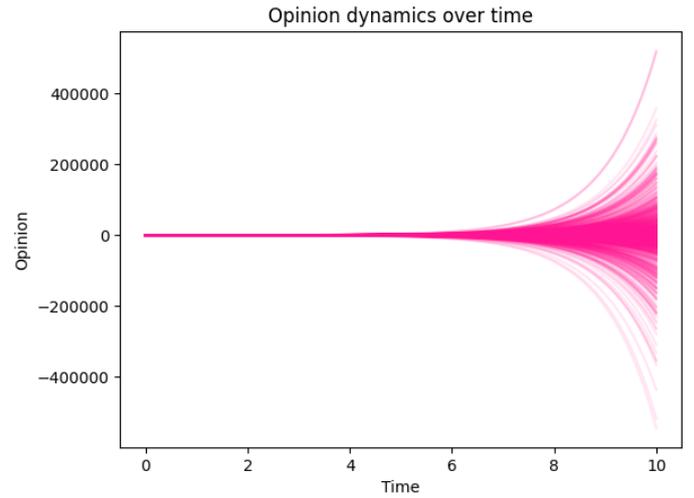

Fig. 5: Calculation result for Opinion dynamics over time $D_{ij}$<0.5, $N = 10000$

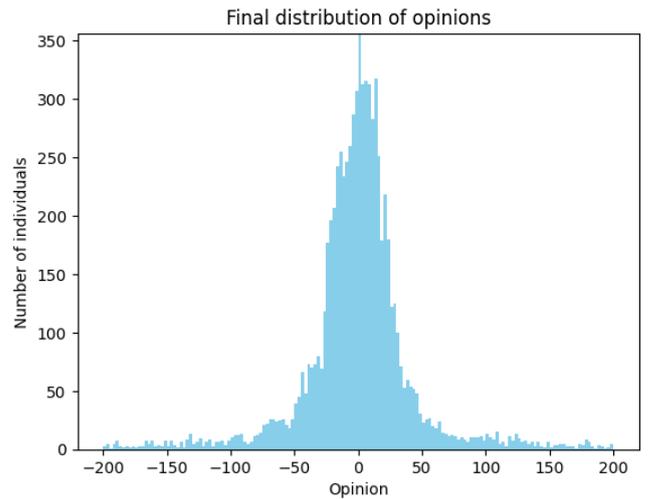

Fig. 6: Calculation result for Final distribution of opinions $D_{ij}$<0.5, $N = 10000$

1. **Differences in considerations between trust $D_{ij} < 0.1$ and**

(1) **Low degree of trust**
Trust $D_{ij} < 0.1$ indicates that the probability that people trust others is very low. As a result, the likelihood that the process of consensus building will be difficult increases. People's opinions rarely converge, reproducing the tendency toward more dispersion of opinions and more extreme opinions.

(2) **Dispersion of opinions**
Compared to the case of trust $D_{ij} < 0.5$, the dispersion of opinions may increase further when trust $D_{ij} < 0.1$. This was observed due to the fact that people are less likely to be influenced by the opinions of others and more likely to hold to their own opinions because of the diminishing trust among people.

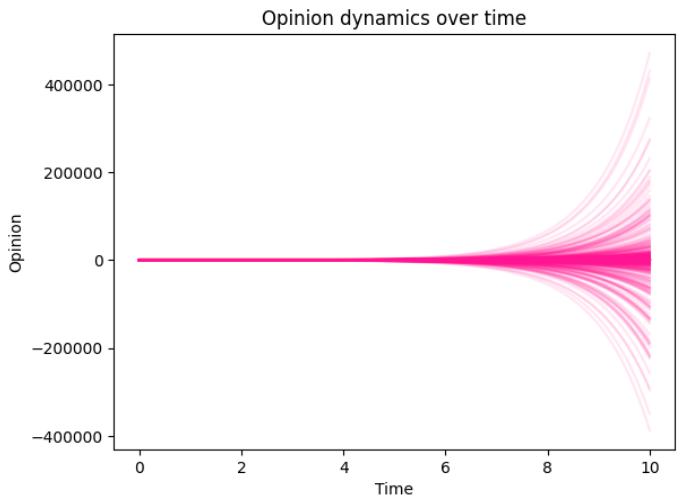

Fig. 7: Calculation result for Opinion dynamics over time $D_{ij}$<0.5, $N$ = 3000

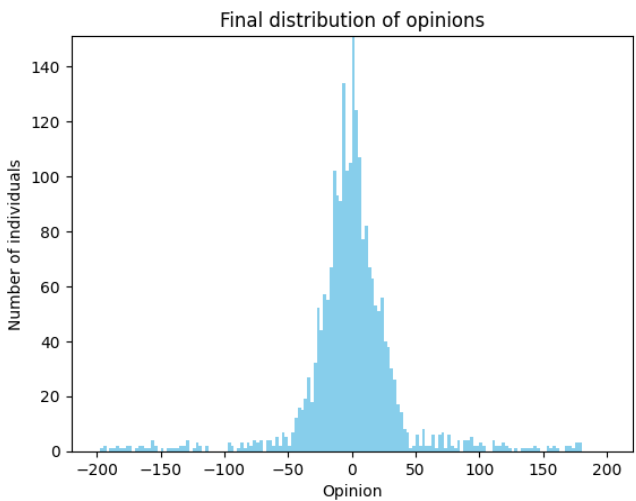

Fig. 8: Calculation result for Final distribution of opinions $D_{ij}$<0.5, $N$ = 3000

## 2. Social Case, Influence of Opinion Formation

When trust $D_{ij} < 0.1$, the effects of information inequality and bias will be stronger in society. With limited reliable sources of information, there will be an increased tendency to rely on a few sources, which can have a significant impact on opinion formation.

The degree to which people form trusting relationships differs greatly between cases of trust $D_{ij} < 0.1$ and trust $D_{ij} < 0.5$. This difference in degree could lead to different trends and outcomes in the process of consensus building and opinion formation.

For $N = 3000$, $Dij < 0.5$, convergence of opinions appears to be more rapid than for $N = 10000$, $Dij < 0.1$. This may be due to the higher probability of trust, which makes people more easily influenced by each other's opinions, and thus tends to cause opinions to converge more quickly.

In the $Dij < 0.5$ condition, opinions converge around a central value, whereas in the $Dij < 0.1$ condition, the distribution of opinions is somewhat more spread out. This indicates that when the probability of trust is low, the convergence of opinions is slower and the distribution of opinions is more spread out.

When considered in a social context, the $Dij < 0.5$ condition would indicate a situation where information is relatively reliable and people are more likely to trust and accept each other's opinions. On the other hand, the $Dij < 0.1$ condition is considered to indicate a situation where information is less reliable and people are skeptical of each other's opinions.

In general, the higher the probability of trust, the more quickly opinions tend to converge and the narrower the distribution of opinions. When the probability of trust is low, convergence of opinions is slower and the distribution of opinions tends to be wider.

Figure 7 and Figure 8,

**Mathematical Formulation for Large Scale $D_{ij}$<0.6**

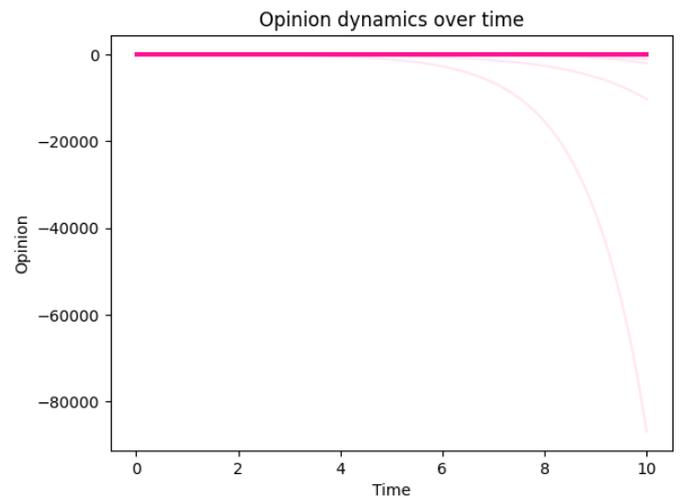

Fig. 9: Calculation result for Opinion dynamics over time $D_{ij}$<0.6, $N$ = 10000

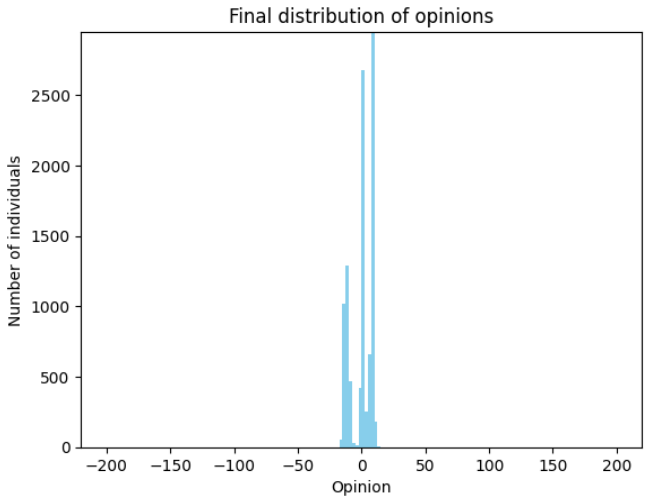

Fig. 10: Calculation result for Final distribution of opinions $D_{ij}$<0.6, $N = 10000$

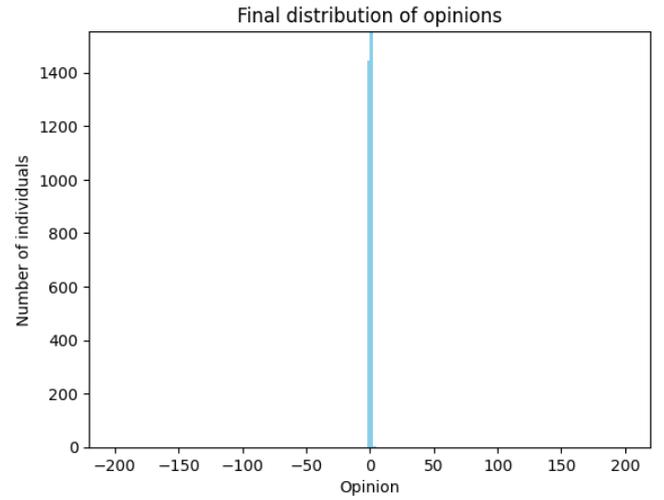

Fig. 12: Calculation result for Final distribution of opinions $D_{ij}$<0.6, $N = 3000$

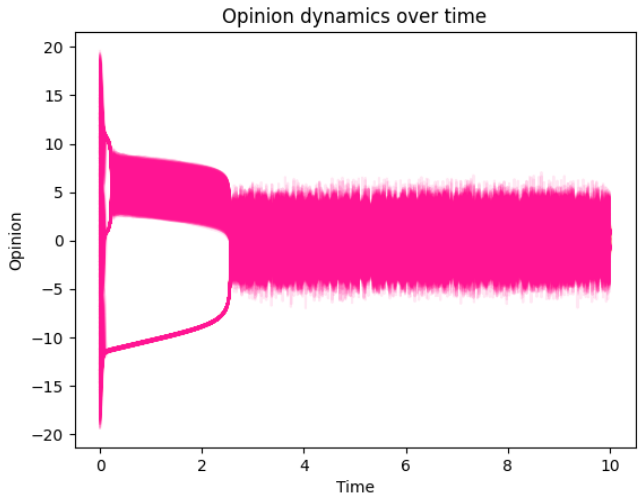

Fig. 11: Calculation result for Opinion dynamics over time $D_{ij}$<0.6, $N = 3000$

### 1. Considerations on the process of consensus building under conditions of trust and distrust

Figure 9 and Figure 10, the *Opinion dynamics over time* graph, opinions tend to decrease with time, suggesting that people are not moving toward consensus. The initial stage of the process is characterized by a small dispersion of opinions, and the dispersion increases over time, suggesting that it is difficult for people to share their opinions.

The *Final distribution of opinions* graph shows that the final distribution of opinions has a peak around 0, with few extreme opinions (very high or low). However, the wide distribution of opinions around 0 can be considered to indicate a state of lack of consensus.

### 2. Cases in society, discussion on opinion formation

Under conditions of trust $Dij < 0.6$, a balance of trust and distrust is maintained among people. However, compared to the aforementioned cases $Dij < 0.1$ and $Dij < 0.5$, clear differences are observed in the dispersion of opinions and the process of consensus formation.

In the $Dij < 0.1$ and $Dij < 0.5$ cases, the dispersion of opinions tended to be larger, while in the $Dij < 0.6$ case, the dispersion of opinions was smaller and more people tended to hold neutral opinions. This is thought to be because the relatively high level of trust among people makes them more susceptible to the opinions of others and more likely to converge on a neutral opinion.

However, even when $Dij < 0.6$, this suggests that people are not in complete agreement. This may indicate that even when trust and distrust are well balanced, sharing opinions and reaching consensus are not easy.

We find that different trust levels produce distinct differences in the process of opinion dispersion and consensus building. At lower trust levels, opinions tend to be more dispersed, while at higher trust levels, opinions tend to converge more neutrally. However, we believe that additional conditions are necessary to reach complete agreement.

Figure 11 and Figure 12,

### 1. Consideration of the process of consensus building under conditions of trust and distrust

As the *Opinion dynamics over time* graph for $N = 3000$ shows, opinions seem to converge rapidly at the beginning.

In particular, as time passes, the distribution of opinions becomes more concentrated, eventually converging around a central value. This rapid convergence indicates a process of consensus building under conditions of trust $Dij < 0.6$, suggesting that opinions quickly converge when trust is high. The *Final distribution of opinions* graph shows that opinions are concentrated near 0. This indicates that there are few extreme opinion values and that most people have a median opinion. This suggests that the majority of people hold similar opinions and are in agreement.

### 2. The difference between the formation of opinions within a society and the aforementioned considerations of $N = 10000$, $Dij < 0.1$, $Dij < 0.5$

Condition of $N = 3000$, $Dij < 0.6$.

This case shows that convergence of opinion is proceeding quickly. This may be due to people being more susceptible to the opinions of others, as the probability of trust is higher under $Dij < 0.6$. In a social context, this condition suggests a situation where information is relatively reliable and people are more likely to trust and accept each other's opinions. In comparison to the aforementioned conditions of $N = 10000$, $Dij < 0.1$, and $Dij < 0.5$: The $Dij < 0.5$ condition shows that opinions converge around a median value, whereas the $Dij < 0.1$ condition shows a slightly wider distribution of opinions. Under the $Dij < 0.1$ condition, the probability of trust is low, so the convergence of opinions is slower and the distribution of opinions tends to be wider. In summary, the higher the probability of trust, the more quickly opinions converge and the narrower the distribution of opinions. When the probability of trust is low, opinions tend to converge more slowly and the distribution of opinions tends to widen.

**Mathematical Formulation for Large Scale** $D_{ij}<0.7$

Figure 13 and Figure 14,

### (1) Consideration as a process of consensus building under conditions of trust and distrust

**Opinion dynamics over time**: As can be seen from the graph, opinions are highly variable over time. This may indicate that it is difficult to reach a certain consensus among people. Opinion fluctuates at different times of the day and may show large fluctuations at certain times of the day while remaining relatively stable at other times of the day. This may suggest that opinions change under the influence of specific events or information.

As can be seen from the **Final distribution of opinions** graph, The final distribution of opinions has a peak near zero, but is dispersed over a very wide range. This indicates that while many people hold neutral opinions, there are also those who hold extreme opinions.

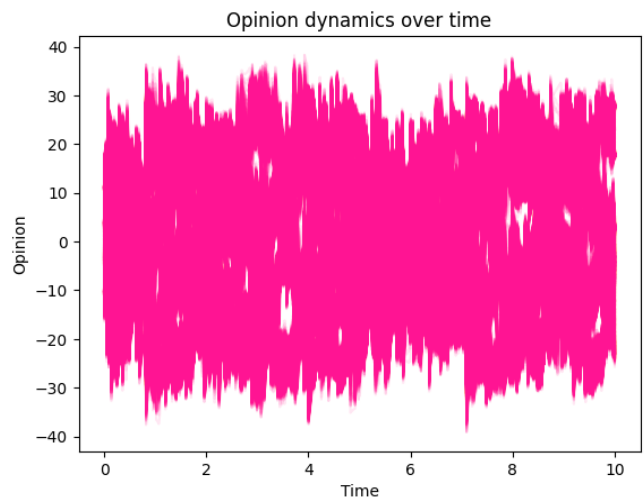

Fig. 13: Calculation result for Opinion dynamics over time $D_{ij}<0.7$

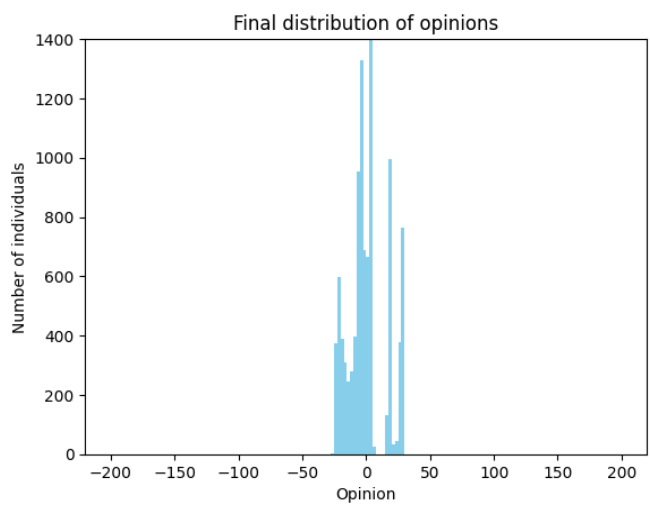

Fig. 14: Calculation result for Final distribution of opinions $D_{ij}<0.7$

### (2) What is the case in society, opinion formation Differences in consideration between the aforementioned trust $Dij < 0.1$, $Dij < 0.5$, and $Dij < 0.6$.

When trust $Dij < 0.7$, most people tend to have a neutral opinion. This suggests that under high levels of trust, people are more likely to be influenced by the opinions of others and more likely to converge on common opinions and values. Compared to $Dij < 0.1$ and $Dij < 0.5$, we see that when $Dij < 0.7$, opinions are less dispersed and more people hold neutral opinions. This indicates that the higher the confidence level, the less extreme the opinion variance. However, even with $Dij < 0.7$, we can confirm that we are not in complete agreement. This indicates that even at high confidence levels, diversity of opinion may be preserved.

The above analysis shows that the level of trust influences the process of opinion formation and consensus building. In particular, the results suggest that under high levels of trust, while opinions fluctuate less, complete agreement is not reached.

**Mathematical Formulation for Large Scale $D_{ij}$<0.8**

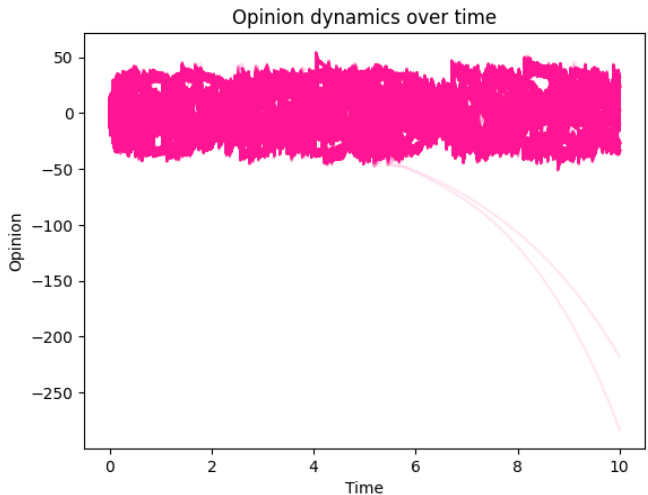

Fig. 15: Calculation result for Opinion dynamics over time $D_{ij}$<0.8

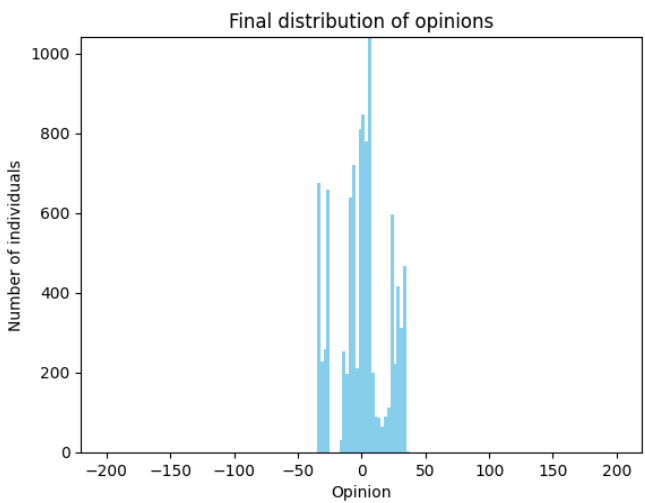

Fig. 16: Calculation result for Final distribution of opinions $D_{ij}$<0.8

Figure 15 and Figure 16,

**(1) Consideration as a process of consensus building under conditions of trust and distrust**

The "Opinion dynamics over time" graph shows that there is a great deal of fluctuation in opinions. This could be due to the possibility of active exchange of information and influence among people. In addition, this fluctuation is particularly large during certain time periods, suggesting that specific events or information may be influencing the formation of opinions.

The "Final distribution of opinions" graph shows that while many people hold neutral opinions, there are also those who hold extreme opinions. This may indicate that diverse opinions are held as information is exchanged and opinions are formed amidst a high level of trust.

**(2) What cases in society, opinion formation Differences in consideration between the aforementioned trust $Dij < 0.1$, $Dij < 0.5$, $Dij < 0.6$, and $Dij < 0.7$**

When the aforementioned values of trust $Dij$ are small, the distribution of opinions is more likely to be expansive, since a small $Dij$ means that trust relationships are likely to be tenuous and individual opinions are likely to be diverse.

Conversely, a large value of $Dij$, e.g., $Dij < 0.7$, is considered to increase the likelihood that the distribution of opinions will be centrally concentrated, since a large $Dij$ is considered to indicate a strong trust relationship and a greater likelihood of convergence of people's opinions.

In conclusion, the strength of the trust relationship ($Dij$ value) may still significantly change the distribution and variability of opinions. By changing the number of trust relationships $Dij$, we can confirm that we can observe fluctuations in the process of opinion forming and consensus building.

**Mathematical Formulation for Large Scale $D_{ij}$<0.9**

Figure 17 and Figure 18,

**(1) Consideration as a process of consensus building under conditions of trust and distrust**

From the graph of the final opinion distribution, it can be read that many individuals hold neutral opinions while others hold extreme opinions. This may indicate that opinions are formed under conditions of high trust as information is actively exchanged.

In the graph of opinion dynamics over time, we can see that opinions fluctuate more during certain time periods, especially in the first part of the graph. This suggests that certain events or information may influence the formation of opinions.

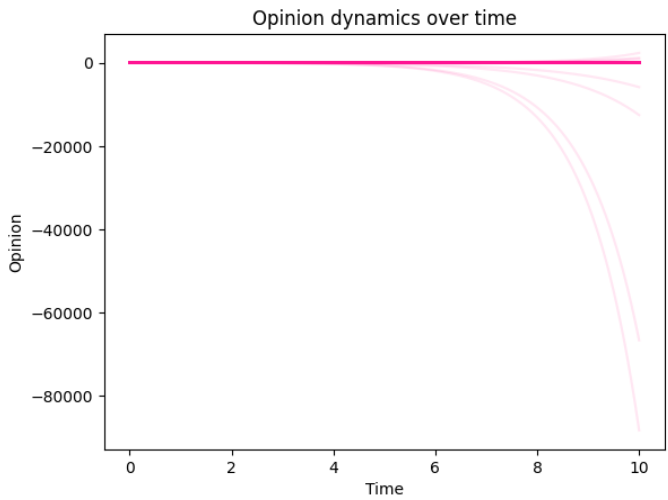

Fig. 17: Calculation result for Opinion dynamics over time $D_{ij}$<0.9

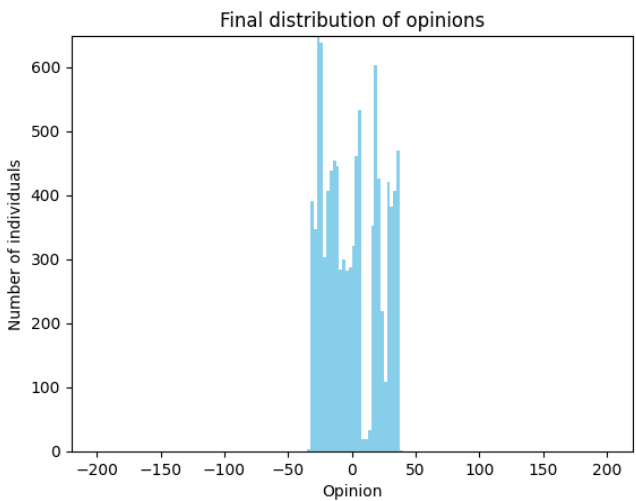

Fig. 18: Calculation result for Final distribution of opinions$D_{ij}$<0.9

**(2) What case in society, opinion formation Differences in consideration between the aforementioned trust $Dij < 0.1$, $Dij < 0.5$, $Dij < 0.6$, $Dij < 0.7$, and $Dij < 0.8$**

Comparing the results under the $Dij < 0.9$ confidence level with the aforementioned considerations from $Dij < 0.1$ to $Dij < 0.8$, we would expect that the distribution of opinions tends to be wider when the confidence level is low (e.g., $Dij < 0.1$). This could indicate that individual opinions are diverse within low trust relationships.

Conversely, at high levels of trust (e.g., $Dij < 0.7$ or $Dij < 0.8$), one would expect to see an increased tendency for the distribution of opinions to be more centrally concentrated. This means that high levels of trust indicate strong trust relationships and the potential for convergence of people's opinions.

The present results for $Dij < 0.9$ confirm that there is a wide range in the distribution of opinions, even within relatively high levels of trust. This could indicate that individual opinions are diverse, even when the level of trust is very high.

**Mathematical Formulation for Large Scale$D_{ij}$<0.99**

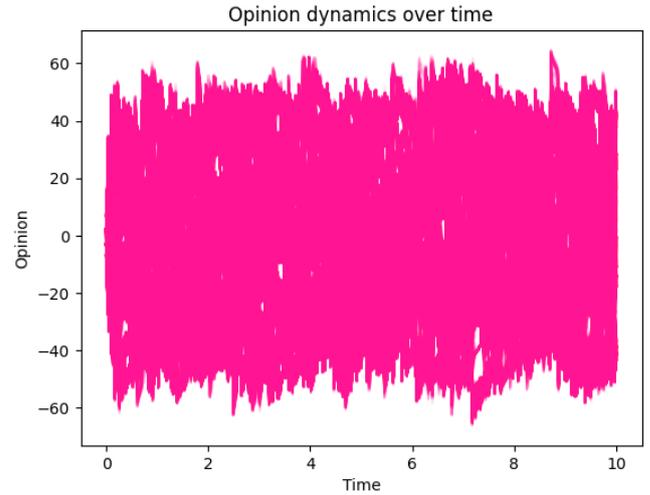

Fig. 19: Calculation result for Opinion dynamics over time $D_{ij}$<0.99, $N = 10000$

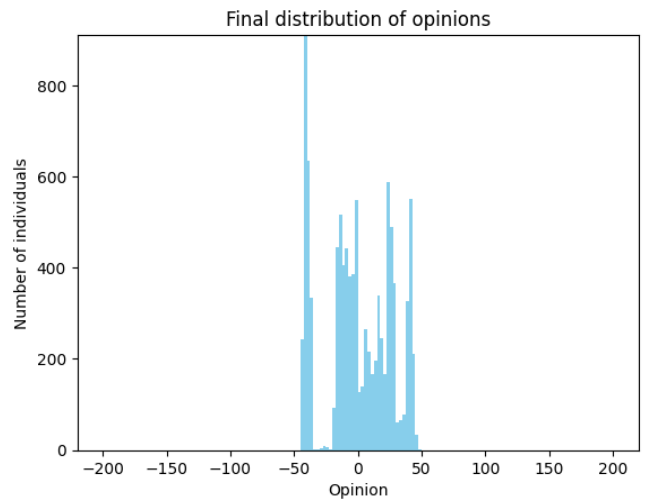

Fig. 20: Calculation result for Final distribution of opinions$D_{ij}$<0.99, $N = 10000$

Figure 19 and Figure 20,

**(1) Consideration as a process of consensus building under conditions of trust and distrust**

The opinion dynamics graph shows how opinion fluctuates over time. In particular, it can be seen that the fluctuation of opinion is large at the beginning of the graph and relatively stable thereafter. This may indicate that information and events in the early stages have a significant impact on the formation of opinions, and that there is a gradual shift toward consensus in the later stages.

The Final Distribution of Opinions graph shows the final distribution of opinions. It shows that many people hold neutral opinions while others hold extreme opinions. This may indicate that opinions tend to form when information is actively exchanged in high trust situations.

**(2) What is the case in society, opinion formation**

Differences in consideration between trust $Dij < 0.1$, $Dij < 0.5$, $Dij < 0.6$, $Dij < 0.7$, $Dij < 0.8$, $Dij < 0.9$: In the case of low confidence (e.g. $Dij < 0.1$): one would expect the distribution of opinions to be more spread out when the confidence level is low. According to the given code, random values are assigned when $Dij$ is less than 0.99, suggesting that opinions may be diverse in low-trust situations.

In high-trust situations (e.g. $Dij < 0.7$ or $Dij < 0.8$): Conversely, one would expect the distribution of opinions to tend to be more centralized in high-trust situations. This means that a high level of trust indicates a strong trust relationship and increases the likelihood of convergence of people's opinions.

Very high trust ($Dij < 0.9$): Even in this case, the distribution of opinions is shown to be spread out. This may indicate that individual opinions are diverse, even at very high confidence levels.

Figure 21 and Figure 22,

**(1) Consideration as a process of consensus building under conditions of trust and distrust**

Final distribution of opinions According to the graph, the distribution of opinions is concentrated around the median. This indicates that many individuals have similar opinions and suggests that opinions converge rapidly under conditions of high trust (consensus formation under $Dij<0.99$).

The Opinion dynamics over time graph shows how opinions fluctuate over time. The graph shows opinion dynamics, indicating that opinions fluctuate within a certain range. However, as time progresses, opinions fluctuate less and less, suggesting that we are gradually moving toward consensus.

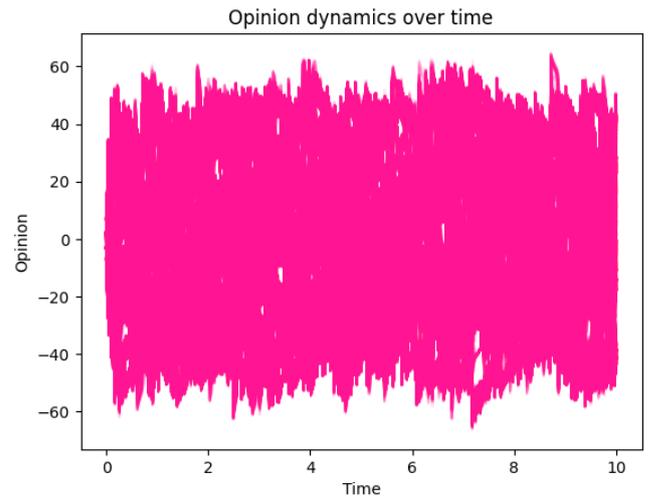

Fig. 21: Calculation result for Opinion dynamics over time $D_{ij}<0.99$, $N = 3000$

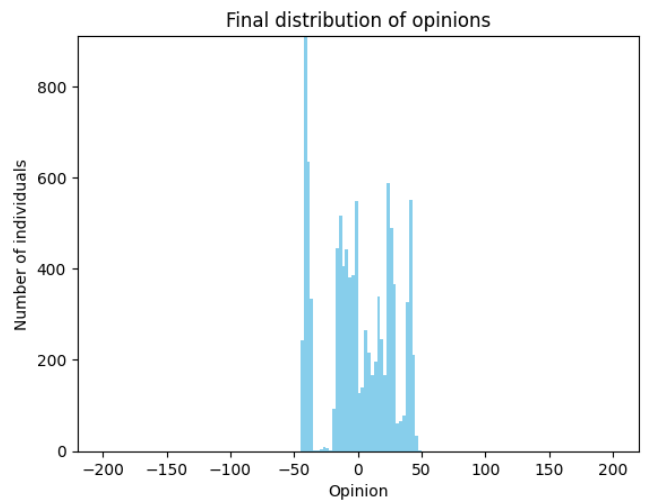

Fig. 22: Calculation result for Final distribution of opinions $D_{ij}<0.99$, $N = 3000$

**(2) What is the case in society, opinion formation**

Differences in consideration between $N = 10000$, $Dij < 0.1$, $Dij < 0.5$, $Dij < 0.6$, and $Dij < 0.99$ mentioned above The $N = 10000$, $Dij < 0.1$ case shows opinion formation under low confidence conditions. In this case, opinions are less likely to converge and the distribution of opinions will tend to be more spread out.

The cases $N = 10000$, $Dij < 0.5$ and $N = 10000$, $Dij < 0.6$ show opinion formation under medium confidence conditions. Under these conditions, we would expect a slightly faster convergence of opinions and a narrower distribution of opinions.

The current case of $N = 3000$, $Dij < 0.99$, shows opinion formation under very high confidence levels. Under these conditions, we observed a rapid convergence of opinions and a very narrow distribution of opinions. This indicates that opinions are more likely to converge under high levels of trust.

## 4. Conclusion

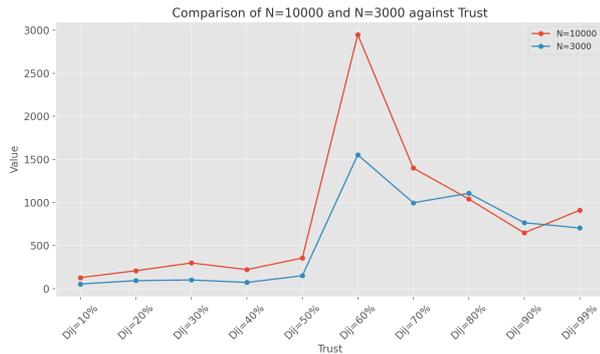

Fig. 23: Comparison of $N = 10000$ and $N = 3000$ against Trust

(1) **Consideration of the phase transition nature of the opinions**

In previous studies, we have examined up to $N = 1600$ as the limit of our computational resources, and now we consider a large-scale case. In both cases, peaks are formed at $Dij = 50\%$ and $Dij = 60\%$. This suggests that once the confidence threshold exceeds a certain point (in this case around $Dij = 50\% - 60\%$), the distribution of social opinion changes abruptly. This is a phenomenon similar to a phase transition, where small changes in a few parameters can cause large fluctuations in results.

(2) **Movement between large and small populations**

We see that the bucket size scores are higher for $n = 10000$ than for $n = 3000$. This indicates that convergence of opinion in large populations may be stronger than in small populations. In large populations, the convergence of opinions at a particular trust threshold may be stronger because of the diversity and different opinions present.

(3) **Trust Thresholds and Distribution of Opinions**

At low ($10\% - 40\%$) or high ($70\% - 90\%$) $Dij$, convergence of opinions tends to be weaker. At low $Dij$, distrust increases and the distribution of opinions is more diffuse. On the other hand, at high $Dij$, different opinions are more likely to persist because most people trust each other.

(4) **Social Context**

This graph illustrates the impact of trust and distrust on opinion formation in social interactions. Specifically, it suggests that when there is a moderate level of trust ($Dij = 50\% - 60\%$), strong convergence and sharing of opinions is more likely to occur within a society. However, at very high levels of trust, this convergence may weaken.

Overall, the results underscore the importance of the role of trust in opinion formation within a society. An appropriate balance of trust may promote convergence and sharing of opinions and strengthen social cohesion.

## Aknowlegement

The author is grateful for discussion with Prof. Serge Galam.This research is supported by Grant-in-Aid for Scientific Research Project FY 2019-2021, Research Project/Area No. 19K04881, "Construction of a new theory of opinion dynamics that can describe the real picture of society by introducing trust and distrust".